\documentclass[twocolumn,showpacs,floatfix,superscriptaddress,amsmath,amssymb,prl]{revtex4}
\usepackage{txfonts}
\usepackage{amssymb}
\usepackage{graphicx}
\usepackage{hyperref}
\usepackage{ulem}
 \usepackage{overpic}
 \usepackage{psfrag}
\newcommand{\be}{\begin{equation}}
\newcommand{\ee}{\end{equation}}

\begin{document}

\title{Fidelity approach to quantum phase transitions:  finite size scaling for quantum Ising model in a transverse field}

\author{Huan-Qiang Zhou, Jian-Hui Zhao and Bo Li}
\affiliation{Centre for Modern Physics and Department of Physics,
Chongqing University, Chongqing 400044, The People's Republic of
China}
\begin{abstract}

We analyze the scaling parameter, extracted from the fidelity for
two different ground states, for the one-dimensional quantum Ising
model in a transverse field near the critical point. It is found
that, in the thermodynamic limit, the scaling parameter is singular,
and the derivative of its logarithmic function with respect to the
transverse field strength is logarithmically divergent at the
critical point. The scaling behavior is confirmed numerically by
performing a finite size scaling analysis for systems of different
sizes, consistent with the conformal invariance at the critical
point. This allows us to extract the correlation length critical
exponent, which turns out to be universal in the sense that the
correlation length critical exponent does not depend on either the
anisotropic parameter or the transverse field strength.
\end{abstract}
\pacs{03.67.-a, 05.70.Fh, 64.60.Ak}

\date{\today}
\maketitle

{\it Introduction.} An emerging picture arises due to latest
advances in quantum information science, which allows us to study
quantum phase transitions (QPTs)~\cite{sachdev} from the ground
state wave functions of many-body systems.  One of the well-studied
aspects is to unveil the possible role of entanglement in
characterizing
QPTs~\cite{preskill,osborne,vidal,entanglement,levin,barnum} (for a
review, see~\cite{amico}). Remarkably, for quantum spin chains, the
von Neumann entropy, as a bipartite entanglement measure, exhibits
qualitatively different behaviors at and off
criticality~\cite{vidal}.

On the other hand, the fidelity, another basic notion of quantum
information science, has attracted a lot of
attention~\cite{zanardi,zjp,more} quite recently. In
Ref.~\cite{zjp}, it has been shown that it may be used to
characterize QPTs, which occur in quantum spin chain, regardless of
what type of internal order is present in quantum many-body states
(either the conventional symmetry-broken orders or exotic QPTs in
matrix product systems~\cite{wolf}). The argument is solely based on
the basic Postulate of Quantum Mechanics on quantum measurements.
Indeed, the basic Postulate of Quantum Mechanics on quantum
measurements implies that two non-orthogonal quantum states are not
reliably distinguishable~\cite{nielsen}. Therefore, any two ground
states must be orthogonal due to the occurrence of orders,
regardless of what type of QPTs. Conversely, the fact that two
ground states are orthogonal implies that they are reliably
distinguishable. Therefore, an order parameter, which may be
constructed systematically in principle~\cite{oshikawa}, exists for
any systems undergoing QPTs. It is the quantitative or qualitative
difference unveiled in order parameters that justifies the
introduction of the notions of irrelevant and relevant information.
To quantify irrelevant and relevant information, the scaling
parameter extracted from the fidelity was introduced to characterize
QPTs. This establishes an intriguing connection between quantum
information theory, QPTs, renormalization group (RG) flows and
condensed matter physics.

The fact that any two different ground states are orthogonal for
continuous QPTs makes it difficult (if not impossible) to extract
physical information solely from ground states themselves.
Conventionally, condensed matter physicists and field theorists
focus on spectra and correlation functions. Therefore, it is
somewhat surprising to see that simply partitioning a system into
two parts and quantifying entanglement between them reveal highly
nontrivial information about QPTs. The intrinsic irreversibility due
to information loss along RG flows may also be revealed solely from
ground states~\cite{vidal,latorre,orus,zbfs}. In the fidelity
approach~\cite{zjp},  it is necessary to put the whole system on a
finite chain, and observe how the fidelity scales with system sizes
as the thermodynamic limit is approached, in order to extract
physical information. The difference between entanglement measures
and the fidelity approach lies in the fact that for the former
different entanglement measures need to be devised to detect
QPTs~\cite{barnum}, whereas the latter succeeds to detect QPTs for
quantum spin chains, regardless of what order is present. The
philosophy behind this is that {\it bipartite entanglement measures
involve partitions and some information is lost due to the fact that
the whole is not simply the sum of the parts, whereas in the
fidelity approach, a system is treated as a whole from the starting
point}.

In this paper, we analyze the scaling parameter, extracted from the
fidelity, for the one-dimensional quantum Ising model in a
transverse field near the critical point.  It is found that, in the
thermodynamic limit, the scaling parameter is singular, and the
derivative of its logarithmic function with respect to the
transverse field strength (the control parameter) is logarithmically
divergent at the critical point. A finite size scaling analysis is
carried out for systems of different sizes, and the scaling behavior
is confirmed numerically, consistent with the conformal invariance
at the critical point. This allows us to extract the correlation
length critical exponent. We have also performed numerics to confirm
the universality, i.e., the correlation length critical exponent
does not depend on either the anisotropic parameter or the
transverse field strength.

{\it The fidelity and the scaling parameter for quantum $XY$ spin
chain.} The quantum $XY$ spin chain is described by the Hamiltonian
\begin{equation}
H= -\sum_{j=-M}^M ( \frac {1+\gamma}{2} \sigma^x_j \sigma^x_{j+1} +
\frac {1-\gamma}{2} \sigma^y_j \sigma^y_{j+1}  + \lambda \sigma^z_j
). \label{HXY}
\end{equation}
Here $\sigma_j^x, \sigma_j^y$, and $\sigma_j^z$  are the Pauli
matrices at the $j$-th lattice site. The parameter $\gamma$ denotes
an anisotropy in the nearest-neighbor spin-spin interaction, whereas
$\lambda$ is an external magnetic field. The Hamiltonian (\ref{HXY})
may be exactly diagonalized~\cite{lieb,pfeuty} as $H=\sum_k
\Lambda_k (c_k^\dagger c_k -1)$, where
$\Lambda_k=\sqrt {(\lambda-\cos(2\pi k/L))^2+\gamma^2 \sin^2 (2\pi
k/L)}$, with $c_k$ and $c_k^\dagger$ denoting free fermionic modes
and $L=2M+1$. The ground state $|\psi\rangle$ is the vacuum of all
fermionic modes defined by $c_k|\psi\rangle=0$, and may be written
as $|\psi\rangle=\prod^{M}_{k=1} (\cos
(\theta_k/2)|0\rangle_k|0\rangle_{-k}-i\sin
(\theta_k/2)|1\rangle_k|1\rangle_{-k}$, where $|0\rangle_k$  and
$|1\rangle_k$ are, respectively, the vacuum and single excitation of
the $k$-th mode, and $\theta_k$ is defined by $\cos \theta_k = (\cos
(2\pi k/L)-\lambda)/\Lambda_k$. Therefore, the fidelity $F$ for two
different ground states $|\psi (\lambda, \gamma)\rangle$ and $|\psi
(\lambda',\gamma)\rangle$ takes the form:
\begin{equation}
F(\lambda,\lambda';\gamma)=\prod^M_{k=1}\cos
\frac{\theta_k-\theta'_k}{2} \label{F},
\end{equation}
where the prime denotes that the corresponding variables take their
values at $\lambda'$. Obviously, $F=1$ if $\lambda=\lambda'$.
Generically, $\cos \frac{\theta_k-\theta'_k}{2} < 1$, therefore the
fidelity (\ref{F}) decays very fast when $\lambda$ separates from
$\lambda'$.

Now let us introduce a fundamental quantity-the scaling parameter
$d(\lambda,\lambda';\gamma)$. For a large but finite $L$, the
fidelity scales as $d^L$, with some scaling parameter $d$ depending
on $\lambda$ and $\lambda'$, due to the symmetry under translation.
Formally, in the thermodynamic limit, $d(\lambda,\lambda')$ may be
defined as
\begin{equation}
\ln d(\lambda,\lambda';\gamma) = \lim_{L \rightarrow \infty} \ln
F(\lambda,\lambda';\gamma) /L.
\end{equation}
The scaling parameter $d(\lambda,\lambda';\gamma)$ enjoys some
properties inherited from the fidelity: (1) symmetry under
interchange $\lambda \longleftrightarrow \lambda'$; (2)
$d(\lambda,\lambda;\gamma)=1$; and (3) $0 \leq
d(\lambda,\lambda';\gamma) \leq 1$.

In the thermodynamic limit, the scaling parameter
$d(\lambda,\lambda';\gamma)$ for the quantum XY model takes the
form:
\begin{equation}
\ln d(\lambda,\lambda';\gamma) = \frac {1}{2\pi} \int ^\pi_0 d\alpha
\ln {\cal F} (\lambda,\lambda';\gamma;\alpha),\label{dforising}
\end{equation}
where
\begin{equation}
{\cal F} (\lambda,\lambda';\gamma;\alpha) = \cos [\vartheta
(\lambda;\gamma;\alpha)-\vartheta(\lambda';\gamma;\alpha)]/2,
\end{equation}
with
\begin{equation}
\cos \vartheta (\lambda;\gamma;\alpha) = (\cos \alpha -
\lambda)/\sqrt {(\cos \alpha -\lambda)^2+\gamma^2 \sin^2 \alpha}.
\end{equation}
A notable feature of the scaling parameter (\ref{dforising}) is
that, besides $d(\lambda,\lambda';\gamma)=
d(\lambda',\lambda;\gamma)$ and $d(\lambda,\lambda;\gamma)=1$,  it
even detects the duality between two phases $\lambda
> 1$ and $\lambda < 1$ for quantum Ising model in a transverse field
($\gamma =1$)~\cite{pfeuty}, since it satisfies
$d(\lambda,\lambda';1)=d(1/\lambda,1/\lambda';1)$.

\begin{figure}[ht]
\begin{overpic}[width=42mm,totalheight=42mm]{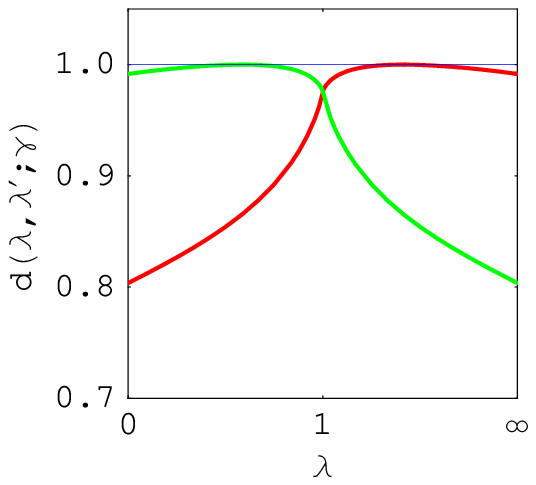}
\put(55,25){$(a)$}
\end{overpic}
\hspace{0in}
\begin{overpic}[width=42mm,totalheight=42mm]{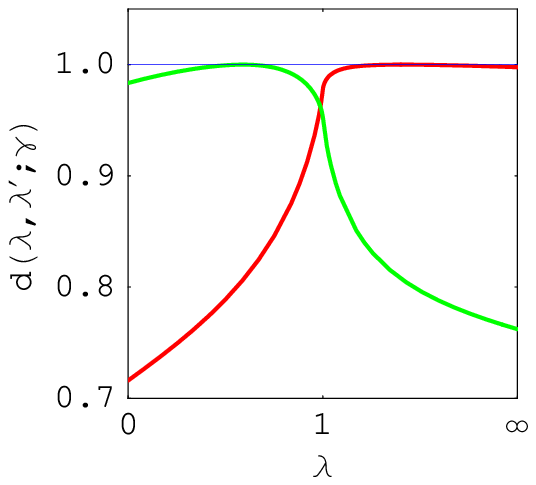}
\put(55,25){$(b)$}
\end{overpic}
\caption{(color online) The scaling parameter
$d(\lambda,\lambda';\gamma)$, extracted from the fidelity for two
ground states $|\psi (\lambda) \rangle$ and $|\psi (\lambda')
\rangle$ of quantum Ising model in a transverse field, is regarded
as a function of $\lambda$ for some fixed values of $\lambda'$ and
$\gamma$. It is continuous but not analytic at $\lambda_c =1$.
($a$): the red line is for $\lambda' =2, \gamma =1$, which touches
the blue line at $\lambda =2$ and the green line is for $\lambda'
=1/2, \gamma =1$, touching the blue line at $\lambda=1/2$. The
mirror symmetry between two curves results from the duality. ($b$):
the green line is for $\lambda' =1/2 , \gamma=1/2$, touching the
blue line at $\lambda =1/2$ and the red line is for $\lambda'
=2,\gamma=1/2$, touching the blue line at $\lambda=2$. No mirror
symmetry for $\gamma \neq 1$.} \label{fig1:1}
\end{figure}

It has been shown~\cite{zjp} that the scaling parameter
$d(\lambda,\lambda';\gamma)$ exhibits a pinch point at ($1,1$),
i.e., an intersection of two singular lines $\lambda =1$ and
$\lambda'=1$, for quantum Ising model in a transverse field
($\gamma=1$). In Fig.\;\ref{fig1:1}, we plot the scaling parameter
$d(\lambda,\lambda';\gamma)$ against $\lambda$ for different values
of $\lambda'$ and $\gamma$. One observes the continuity, as it
should be for continuous QPTs.

Let us focus on the quantum Ising universality class with the
critical line $\gamma\neq 0$ and $\lambda_c=1$. There is only one
(second-order) critical point $\lambda_c=1$ separating two gapful
phases: (spin reversal) $Z_2$ symmetry-breaking and symmetric
phases.  The order parameter, i.e., magnetization $\langle \sigma ^x
\rangle$ is non-zero for  $\lambda < 1$, and otherwise zero. At the
critical point, the correlation length $\xiup \sim |\lambda
-\lambda_c |^{\nu}$ with $\nu = 1$~\cite{pfeuty}. Our purpose is to
extract the correlation length critical exponent by performing a
finite size scaling analysis for $d(\lambda,\lambda';\gamma)$.

\begin{figure}[t]
   \begin{center}
      \begin{overpic}[width=85mm,totalheight=55mm]{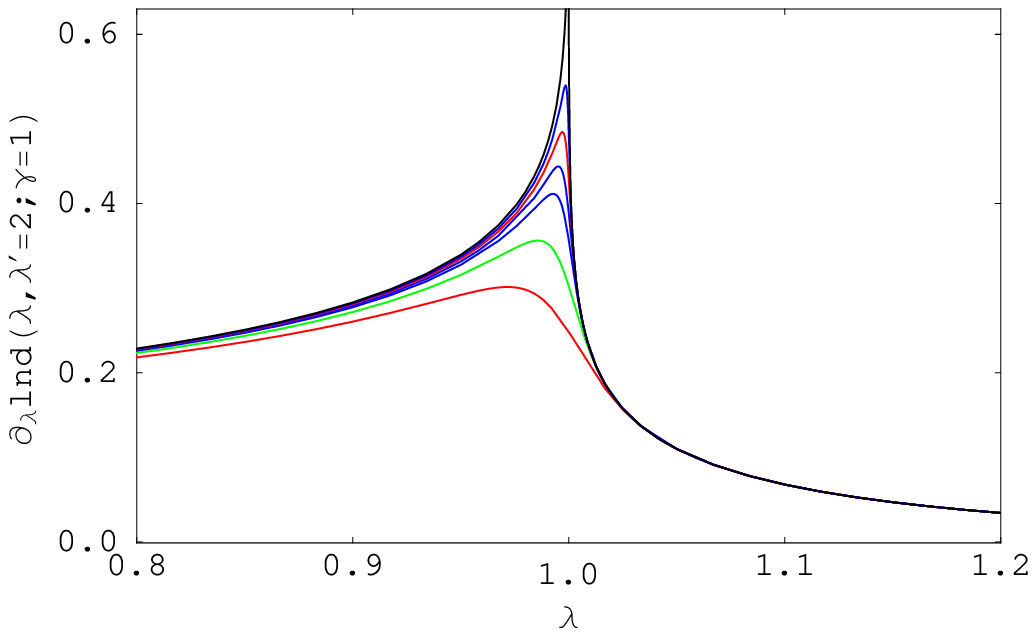}
        \put(53,22){
          \includegraphics[width=32mm,totalheight=28mm] {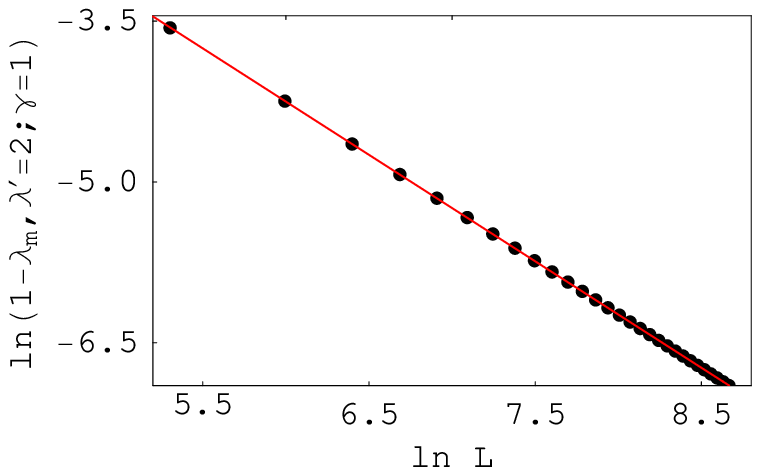}}
    \end{overpic}
\end{center}
  \caption{(color online) Main: the logarithmic divergence near the critical
  point $\lambda_c=1$ is analyzed. This is achieved by considering
  $\partial_{\lambda}{\ln d(\lambda,\lambda'=2; \gamma=1)}$ as the function of the
  transverse field strength $\lambda$. The curves shown correspond to different lattice sizes $L=201,401,1201,2001,4001,\infty$.
  The maximum gets more pronounced, with the system size increasing.
  Inset: the position of maximum approaches the critical
  point $\lambda_c =1$ as $\lambda_m \sim 1-5.52233 L^{-0.99321}$.}
  \label{fig2:2}
\end{figure}

{\it Finite size scaling.} In order to quantify the drastic change
of the ground state wave functions when the system undergoes a QPT
at the critical point $\lambda_c =1$, we evaluate the derivative of
$\ln d(\lambda,\lambda';\gamma)$ with respect to $\lambda$. In the
thermodynamic limit, $\ln d(\lambda,\lambda';\gamma)$ is
logarithmically divergent at the critical point $\lambda_c = 1$:
\begin{equation}
\frac {\partial {\ln d(\lambda,\lambda';\gamma)}}{\partial \lambda}
= k_1 \ln |\lambda -\lambda_c| + {\rm constant} \label{infinite},
\end{equation}
where the prefactor $k_1$ is non-universal in the sense that it
depends on $\lambda'$ and $\gamma$.  The numerical results  are
plotted in Fig.\;{\ref{fig2:2}} for $\lambda'=2$ and $\gamma=1$. The
least square method  yields $k_1 \approx - 0.079742$. For systems of
finite sizes $L$'s, there are no divergence in the derivatives of
$\ln d(\lambda,\lambda';\gamma)$ with respect to $\lambda$, since
the second-order QPT only occurs in the thermodynamic limit.
Instead, some pronounced peaks occur at the so-called quasi-critical
points $\lambda_m$ that approach the critical value as $\lambda_m
\sim 1-5.52233L^{-0.99321}$, with the peak values logarithmically
diverging with increasing system size $L$,
\begin{equation}
\frac {\partial {\ln d(\lambda,\lambda';\gamma)}}{\partial
\lambda}\Big|_{\lambda =\lambda_m} = k_2 \ln L + {\rm constant}
\label{finite},
\end{equation}
where the non-universal prefactor $k_2$ takes the value $k_2 \approx
0.079773$. The scaling ansatz in the systems exhibiting logarithmic
divergences~\cite{barber} requires that the absolute value of the
ratio $k_1/k_2$ is the correlation length critical exponent $\nu$.
In this case, $|k_1/k_2|\sim 0.999613$, very close to the exact
value 1.

\begin{figure}[ht]
  \begin{center}
  \begin{overpic}[width=85mm,totalheight=55mm]{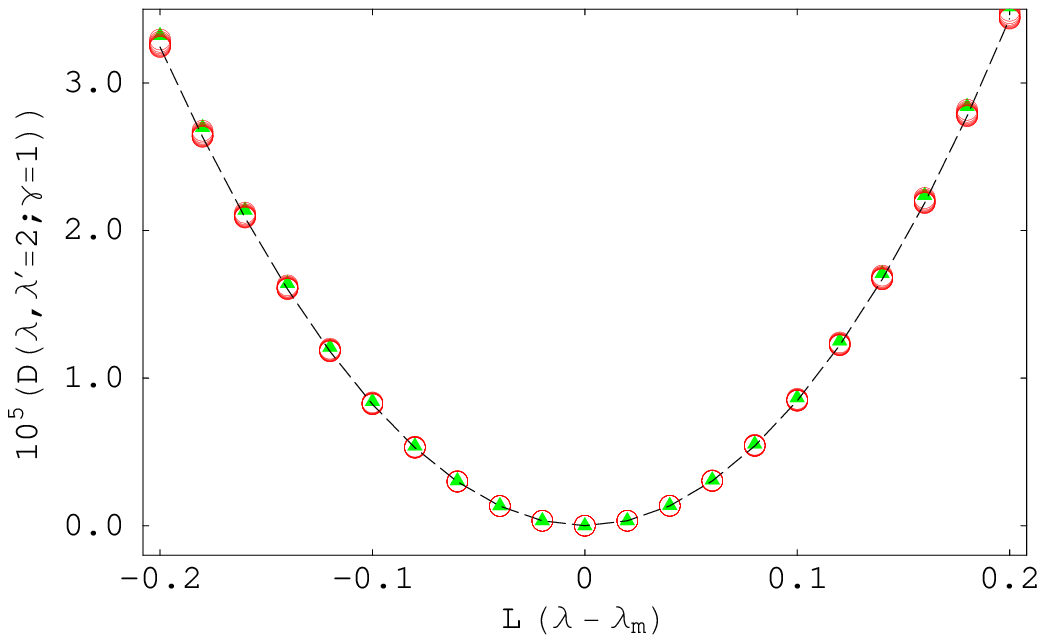}
  \put(28,26){
 \includegraphics[width=38mm,totalheight=26mm] {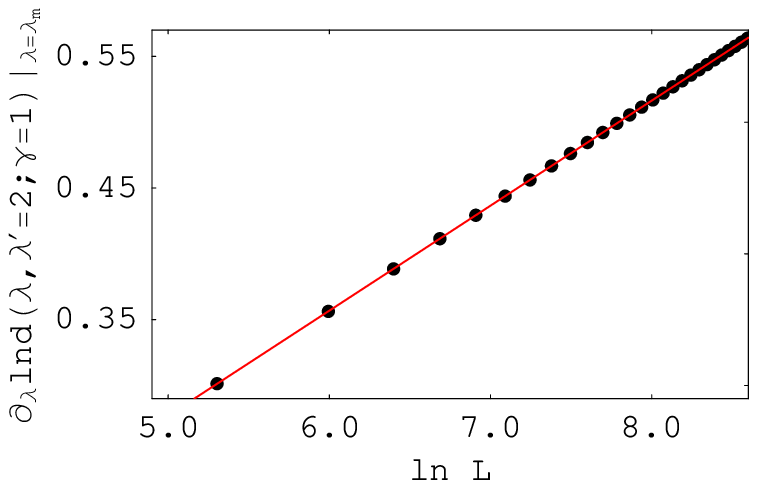}}
\end{overpic}
 \end{center}
\caption{(color online)
 Main: a finite size scaling analysis is carried out for a quantity defined as $D(\lambda,\lambda';\gamma) = 1-\exp[\partial _{\lambda} \ln
  d(\lambda,\lambda';\gamma)- \partial _{\lambda} \ln d(\lambda,\lambda';\gamma)|_{\lambda = \lambda_m}]$.
   According to the finite size scaling ansatz in the case of
logarithmic divergences, one expects that
$D(\lambda,\lambda';\gamma)$ is a function of $L (\lambda
-\lambda_m)$.  Indeed, all the data from $L=801$ up to $L=4001$
collapse on a single curve. This shows that the system at the
critical point is scaling invariant (and thus comformally invariant)
and that the correlation length critical exponent $\nu$ is 1. Inset:
the peak value of $\partial_{\lambda}{\ln d(\lambda,\lambda'=2;
\gamma=1)}$ at $\lambda_m$ diverges as the system size increases,
leading to $k_2 \approx 0.0797726$.}
  \label{fig3:3}
\end{figure}

In the case of logarithmic divergences, a proper scaling ansatz has
been addressed in Ref.~\cite{barber}. Taking into account the
distance of the maximum of $\partial_{\lambda} \ln
d(\lambda,\lambda';\gamma)$ from the critical point, we choose to
plot $1-\exp[\partial _{\lambda} \ln d(\lambda,\lambda';\gamma)-
\partial _{\lambda} \ln d(\lambda,\lambda';\gamma)|_{\lambda =
\lambda_m}]$ as a function of $L(\lambda-\lambda_m)$ for different
system sizes $L$'s. All the data for different $L$'s collapse onto a
single curve. The numerical results for the size ranging form
$L=201$ up to $L=4001$ are plotted in Fig.\;{\ref{fig3:3}}. All
these indicates that the system is scaling invariant, i.e., $\xiup/
L = \xiup'/ L'$ (and thus conformally invariant), and that the
correlation length critical exponent $\nu =1$.

\begin{figure}[ht]
 \begin{center}
 \begin{overpic}[width=85mm,totalheight=55mm]{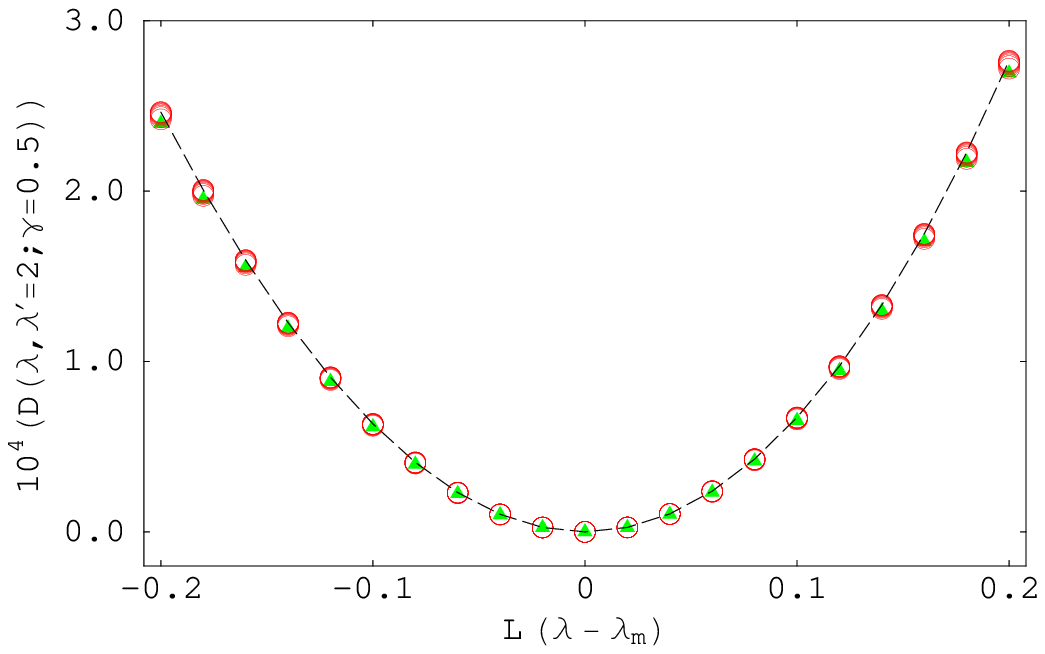}
       \put(30,25){
        \includegraphics[width=38mm,totalheight=26mm] {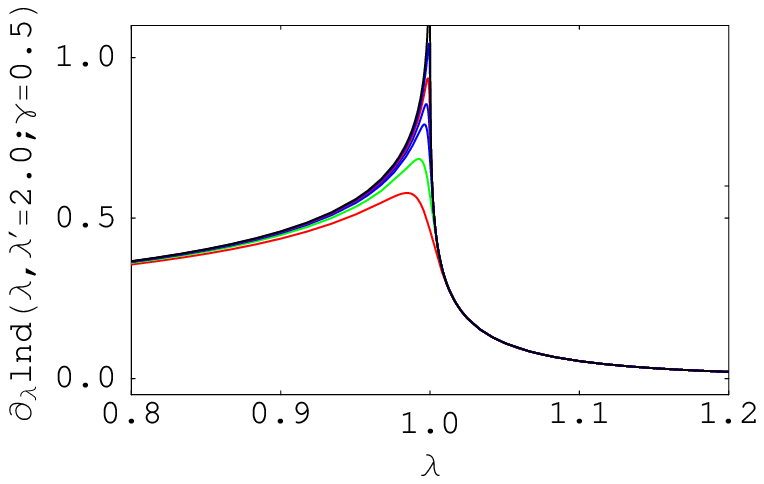}}
    \end{overpic}
\end{center}
\caption{(color online) The universality hypothesis for the scaling
parameter extracted from for the quantum Ising model in a transverse
field is checked against different values of $\gamma$ and
$\lambda'$. Main: in this case we have chosen $\gamma=1/2$ and $L$
ranging from 2801 up to 6001. All the data collapse, consistent with
the fact that the correlation length critical exponent $\nu$ is 1.
The inset shows that the derivative of the logarithmic function of
the scaling parameter $d(\lambda,\lambda';\gamma)$ with respect to
$\lambda$ for $\lambda'=2$ and $\gamma=1/2$ is logarithmically
divergent at $\lambda_c=1$, with $\lambda_m \sim 1- 3.23906
L^{-1.01135}$.
 }
  \label{fig4:4}
\end{figure}

{\it Universality.} As is well known, the quantum XY chain belongs
to the same quantum Ising universality class for non-zero $\gamma$,
with the same critical exponents. To confirm the universality, we
need to check the scaling behaviors for different values of
$\gamma$.  For $\lambda'=2$ and $\gamma=1/2$, in the thermodynamic
limit, it takes the form (\ref{infinite}) with $k_1 \approx
-0.157162$, as long as the control parameter is close to the
critical point, whereas for a system of finite size, it takes the
form (\ref{finite}) with $k_2 \approx 0.157176$. Thus, the absolute
value of the ratio $k_1/k_2$ is $|k_1/k_2| =0.999910$.
Fig.\;{\ref{fig4:4}} shows that all the data for different $L$'s
collapse onto a single curve. We also plot the derivative of the
logarithmic function of the scaling parameter
$d(\lambda,\lambda';\gamma)$ with respect to $\lambda$ for
$\lambda'=2$ and $\gamma=1/2$ (see the inset in
Fig.\;{\ref{fig4:4}}).  All the above results show that the critical
exponent $\nu=1$.

\begin{figure}[ht]
 \begin{center}
 \begin{overpic}[width=85mm,totalheight=55mm]{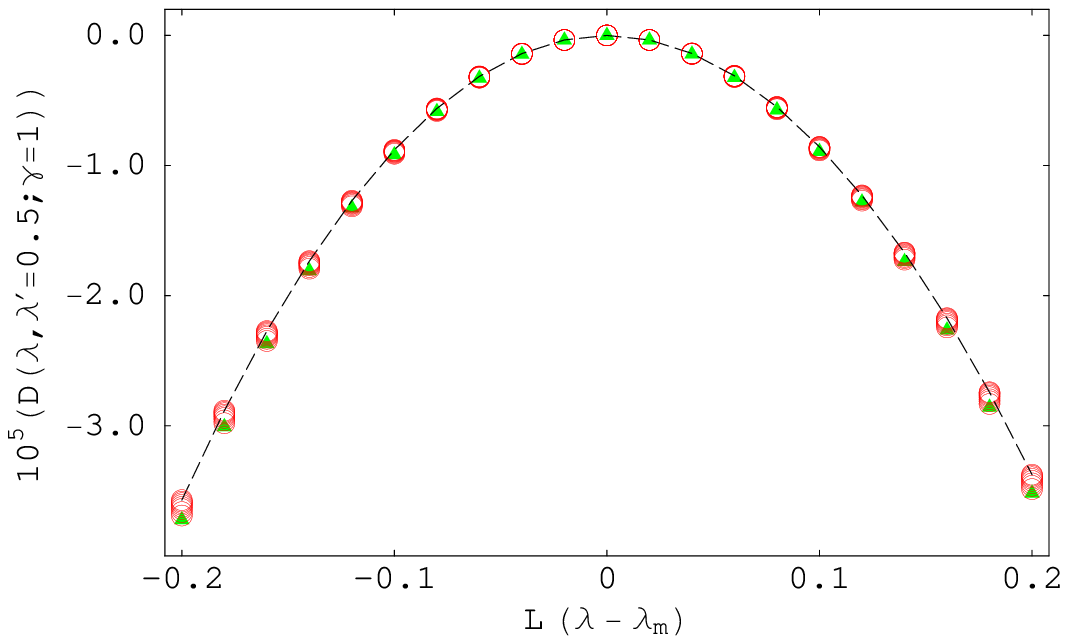}
       \put(30,10){
        \includegraphics[width=38mm,totalheight=26mm] {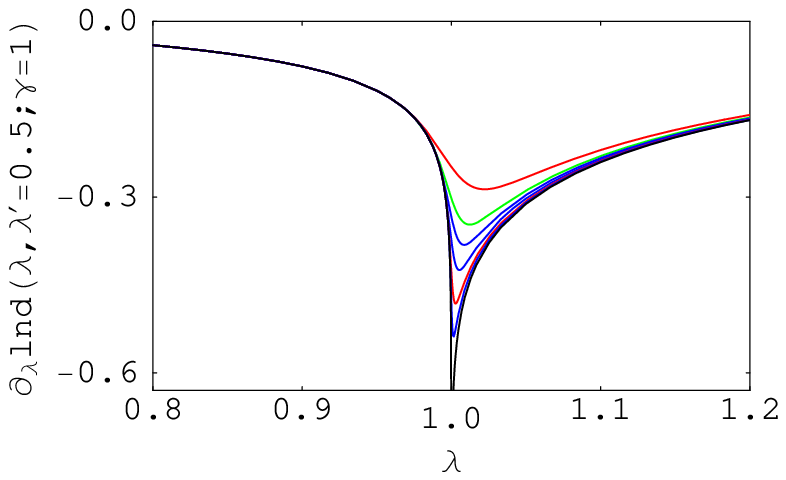}}
    \end{overpic}
\end{center}
\caption{(color online) The universality for the scaling parameter
is checked against different values of $\gamma$ and $\lambda'$.
Main: in this case we have chosen $\gamma=1$ and $L$ ranging from
2801 up to 6001. Consistent with the universality hypothesis for the
quantum Ising model in a transverse field, all the data collapse,
indicating that the correlation length critical exponent $\nu$ is 1.
The inset shows that the derivative of the logarithmic function of
the scaling parameter $d(\lambda,\lambda';\gamma)$ with respect to
$\lambda$ for $\lambda'=1/2$ and $\gamma=1$ is logarithmically
divergent at $\lambda_c =1$, with $\lambda_m \sim 1+ 3.50186
L^{-0.94107}$.
 }
  \label{fig5:5}
\end{figure}

Besides $\gamma$, the scaling parameter $d(\lambda,\lambda';\gamma)$
also depends on the control parameter $\lambda'$.  For
$\lambda'=1/2$ and $\gamma=1$, in the thermodynamic limit, the
derivative of the logarithmic function of the scaling parameter
$d(\lambda,\lambda';\gamma)$ with respect to $\lambda$ still takes
the form (\ref{infinite}), with $k_1 \approx 0.083005$, as long as
the control parameter is close to the critical point, whereas for a
system of finite size, it takes the form (\ref{finite}), with $k_2
\approx - 0.083007$. Thus, the absolute value of the ratio $k_2/k_1$
is $|k_2/k_1| = 0.999975$ , again close to the exact value 1.
Similarly, all the data for different $L$'s collapse onto a single
curve, as shown in Fig.\;{\ref{fig5:5}}.  In the inset, we plot the
derivative of the logarithmic function of the scaling parameter
$d(\lambda,\lambda';\gamma)$ with respect to $\lambda$ for
$\lambda'=1/2$ and $\gamma=1$. Therefore, we have demonstrated that
the universality hypothesis is valid for the scaling parameter
$d(\lambda,\lambda';\gamma)$.

{\it Discussions and conclusions.} As a basic notion of quantum
information science, fidelity may be used to detect QPTs in
condensed matter systems. Remarkably, an intimate connection exists
between RG flows, QPTs and the scaling parameter which may be
extracted from the fidelity~\cite{zjp}. The scaling parameter is
well defined in the thermodynamic limit, in sharp contrast to the
fidelity itself that always vanishes for continuous QPTs. Different
from a bipartite entanglement measure, the fidelity approach does
not involve the partition of the whole system into different parts,
and the system is treated as a {\it whole} from the starting point.
{\it In some sense, such a difference may be counted as the
contribution from multipartite entanglement}. Therefore one may
expect that the fidelity approach possesses significant advantage
over the conventional bipartite entanglement approach~\cite{zhou}.

Another feature worth to be mentioned is that fidelity is simple to
be evaluated in the matrix product state (MPS)
representation~\cite{zjp}. On the other hand, many efficient
numerical algorithms are now available due to the latest
developments in classical simulation of quantum
systems~\cite{porras, vidal1, mcculloch}. This makes it practical to
determine all information including stable and unstable fixed points
along RG flows~\cite{zjp}, and to extract critical exponents from
the scaling parameter, as shown for the quantum Ising model in a
transverse field. In this regard, algorithms for periodic boundary
conditions~\cite{porras} and infinite systems~\cite{vidal1} are
powerful enough to extract meaningful information for critical
systems.

In summary, we have performed a finite size scaling analysis for the
scaling parameter, whose analytical expression has been extracted
from the fidelity for two ground states corresponding to different
values of the control parameter for the one-dimensional quantum
Ising model in a transverse field near the critical point. In the
thermodynamic limit, the logarithmical divergence of the derivative
of the scaling parameter with respect to the transverse field
strength is demonstrated numerically, consistent with the conformal
invariance at the critical point. This makes it possible to extract
the correlation length critical exponent. The latter turns out to be
universal, in the sense that the correlation length critical
exponent thus extracted does not depend on either the anisotropic
parameter $\gamma$ or the transverse field strength $\lambda$.

We thank Sam Young Cho, John Fjaerestad and Jon Links for helpful
discussions and comments.

\end{document}